\documentclass[a4paper]{jpconf}
\usepackage{graphicx}  
\usepackage{dcolumn}   
\usepackage{bm}        
\usepackage[english]{babel}
\usepackage{amsfonts,amsmath,amssymb,mathrsfs}
\usepackage{epstopdf}
\usepackage{subfigure}
\usepackage{color}

\def\a{\alpha}
\def\r{\rho}
\def\s{\sigma}
\def\t{\tau}
\def\m{\mu}
\def\n{\nu}
\def\k{\kappa}
\def\th{\theta}
\def\g{\gamma}\def\G{\Gamma}
\def\L{\Lambda}\def\l{\lambda}
\def\D{\Delta}
\def\la{\langle}
\def\ra{\rangle}
\def\o{\omega}\def\O{\Omega}
\def\d{\delta}
\def\p{\partial}

\def\half{\textstyle{\frac{1}{2}}}

\def\bdoc{\begin{document}}
\def\edoc{\end{document}}

\def\beq{\begin{equation}}
\def\eeq{\end{equation}}
\def\bea{\begin{eqnarray}}
\def\eea{\end{eqnarray}}
\def\ben{\begin{enumerate}}
\def\een{\end{enumerate}}
\def\la{\langle}
\def\ra{\rangle}
\def\a{\alpha}
\def\b{\beta}
\def\g{\gamma}\def\G{\Gamma}
\def\d{\delta}\def\D{\Delta}
\def\e{\epsilon}

\def\th{\theta}
\def\k{\kappa}
\def\l{\lambda}
\def\m{\mu}
\def\n{\nu}
\def\o{\omega}
\def\p{\pi}
\def\r{\rho}
\def\s{\sigma}
\def\t{\tau}
\def\L{{\cal L}}
\def\S{\Sigma }
\def\gsim{\; \raisebox{-.8ex}{$\stackrel{\textstyle >}{\sim}$}\;}
\def\lsim{\; \raisebox{-.8ex}{$\stackrel{\textstyle <}{\sim}$}\;}
\def\gtrsim{\gsim}
\def\lessim{\lsim}
\def\loc{{\rm local}}
\def\vm{v_{\rm max}}
\def\bh{\bar{h}}
\def\del{\partial}
\def\nab{\nabla}
\def\half{{\textstyle{\frac{1}{2}}}}
\def\fourth{{\textstyle{\frac{1}{4}}}}

\def\bD{{\bf D}}
\def\bE{{\bf E}}
\def\bF{{\bf F}}
\def\bB{{\bf B}}
\def\bP{{\bf P}}
\def\bV{{\bf v}}
\def\bv{{\bf v}}
\def\bx{{\bf x}}
\def\by{{\bf y}}
\def\bz{{\bf z}}
\def\ba{{\bf a}}
\def\bd{{\bf d}}
\def\bs{{\bf s}}
\def\bn{{\bf n}}
\def\bp{{\bf p}}

\def\O{\Omega}

\def\br{{\bf r}}
\def\bnab{{\bf \nab}}

\def\tE{\tilde{E}}
\def\tL{\tilde{L}}

\begin{document}
\title{Destroying black holes with test bodies}
\author{Ted Jacobson$^*$ and Thomas P. Sotiriou$^{\dagger}$}

\address{$^*$Center for Fundamental Physics,  University of Maryland, College Park, MD 20742-4111, USA\\
$^\dagger$Department of Applied Mathematics and Theoretical Physics, Centre for  
Mathematical Sciences, University of Cambridge, Wilberforce Road,  
Cambridge, CB3 0WA, UK}

\ead{jacobson@umd.edu; T.Sotiriou@damtp.cam.ac.uk}

\begin{abstract}

If a black hole can accrete a body whose spin or charge 
would send the black hole parameters over the extremal limit, 
then a naked singularity would presumably form,
in violation of the cosmic censorship conjecture. 
We review some previous results on testing cosmic censorship
in this way using the test body approximation, focusing 
mostly on the case of neutral black holes.  Under certain 
conditions a black hole can indeed be over-spun or over-charged 
in this approximation, hence radiative and self-force effects must
be taken into account to further test cosmic censorship.

\end{abstract}

\section{Black holes and naked singularities}

Penrose and Hawking have shown that singularities are inevitable in gravitational collapse \cite{Penrose:1964wq,Hawking:1969sw}. These results do not rely on specific solutions of Einstein's equations or special symmetry requirements. Therefore, spacetime singularities are physically relevant and not simply mathematical peculiarities of special solutions. While the results do not say anything about the precise nature of the singularities, they do indicate
the breakdown of general relativity. Einstein's theory is unable to predict the outcome of events in the vicinity of the singularity, and unusual phenomena could take place there. Visible singularities could provide exciting access to the so far unobserved physics of quantum gravity.

In the case of a black hole, however, the event horizon hides the 
singularity. Although this is a lost opportunity for quantum gravity 
phenomenology, it is also a blessing,
if all we wish to understand is what can be seen
from outside without having to grapple with the physics of a singularity.
But does an event horizon always hide
any singularities that form from a collapsing object?
As Roger Penrose first put it in 1969~\cite{Penrose:1969pc}, 
``does there exist a Òcosmic
censorÓ who forbids the appearance of naked singularities, 
clothing each one in an absolute event horizon?" The conjecture that
such a censor indeed exists is called the ``cosmic censorship conjecture".

The physics of event horizons, unlike that of singularities,  
should be well described by classical general relativity, 
so cosmic censorship can be tested and possibly proven without going
beyond known physics. 
 Evidence 
from several directions suggests that any singularity arising to the 
future of generic (not infinitely finely tuned)
non-singular initial conditions, may indeed always be hidden behind a black hole 
event horizon~\cite{Wald:1997wa,Penrose:1999vj}. However, we are far from
having a definitive proof.

Much of the evidence in favor of cosmic censorship comes from
the failure of attempts to violate it in thought experiments. Given the difficulty
of a direct attack on the general question, this strategy
continues to offer an attractive approach to the problem. 
Even in the context of simplifying approximations, 
such thought experiments can uncover mechanisms
tending to uphold censorship, and they can produce
scenarios where censorship would be violated if the 
approximations
were valid. In the latter case, they focus our attention on
those ``dangerous" scenarios and on the limits of validity 
of the approximation scheme. This type of approach is what
will be discussed here\footnote{This paper is based partially on a talk given by T.P.S. at the 1st Mediterranean conference on Classical and Quantum gravity following the lines of Ref.~\cite{Jacobson:2009kt} and partially on an essay prepared for the ``FQXi essay contest: 
What is
Ultimately Possible in Physics?'' \cite{Jacobson:2009kt2}.}.

\section{ ``Destroying'' a black hole}

In general relativity, the spacetime in the vicinity of the black hole is described by the Kerr--Newman (K-N) metric which contains 
three parameters:
the mass $M$,  the spin angular momentum $J$, and  the electric
charge $Q$.
The K-N metric describes a black hole as long as the mass is sufficiently large compared to a combination of the charge and angular momentum, 
$M^2 \ge a^2+Q^2$, where $a=J/M$. (We adopt units with Newton's constant $G$ 
and the speed of light $c$ both set equal to unity.) 
The case where $M^2 = a^2 + Q^2$ is called an extremal black hole,
while for $M^2 < a^2+Q^2$ there is no event horizon and the K-N metric actually describes a naked singularity. Therefore, it would naively seem that to create a naked singularity all one need do is to start with a black hole and toss in matter with enough angular momentum or charge so as to drive its parameters beyond the extremal limit, leaving it no option but to expose the singularity.

On further thought there are some subtleties involved in such a scenario:
\begin{enumerate}
\item  The K-N metric describes stationary configurations,
so the proposed strategy can work as stated only if, after having absorbed the matter, 
the system settles down to a stationary configuration containing all the mass, angular
momentum and charge, i.e. without having shed the excess angular momentum
or charge in the settling down process. Such an outcome is by no means
guaranteed. Indeed, it seems rather unlikely, given the evidence that
the trans-extremal K-N metric is unstable~\cite{Dotti:2006gc,Dotti:2008js}. 
If such instability occurs (the uncertainty lies 
in the proper boundary conditions at the singularity), the system may well 
shed sufficient angular momentum and/or charge, or
not settle down
to such a metric, and at present nobody knows what it would do. 
What this means is that to demonstrate the creation of a naked singularity one
would have to follow the evolution further than the initial ``absorption" of the 
extra matter. So the possibility of initially 
overspinning or overcharging an initial black hole configuration
can only be taken as an indication that cosmic censorship {\it might} fail.
 
\item The notion of ``exposing" the singularity may be 
inappropriate, since the singularity inside a perturbed charged or rotating stationary
black hole cannot send signals to any point, even those interior to the horizon.
(Here we assume that the Cauchy horizon inside the black 
hole is indeed unstable, as evidence indicates~\cite{Cauchy}.)
That is, it has no nonsingular future. 
Hence it is not so clear that, if a horizon could be 
``destroyed", the result would be to expose the singularity that would have been 
there had the horizon {\it not} been destroyed.  It might produce
a wholly different singularity.  Nevertheless, either way 
the process would violate cosmic censorship.
\end{enumerate}
Setting these issues aside for the present, 
we focus here just on the question of whether 
a black hole can initially absorb sufficient 
angular momentum or charge to send its parameters
over the extremal limit.

Given the non-linear nature of Einstein's equations, following
the evolution of a black hole exactly is a very difficult problem that presumably requires 
numerical solution of the Einstein equation. Therefore, most studies of cosmic
censorship to date 
either imposed symmetry conditions or
have been carried out in the simple framework of a test-body 
moving on the black hole spacetime.
The test-body approximation imposes the conditions
\beq
\label{small}
\delta E\ll M, \quad  \delta J\ll M^2, \quad \delta Q\ll M,
\eeq
where $\delta E$, $\delta J$, and $\delta Q$  denote the energy, angular momentum and charge 
of the body. These conditions seem sufficient to ensure that
the influence of the test body can be treated as a small perturbation
on the background spacetime. (In particular, we need not assume
that  $\delta J\ll J$ or $\delta Q\ll Q$, since addition of angular momentum
to a nonspinning black hole, or charge to a neutral black hole, 
can perfectly well be a small perturbation.)
However, they certainly do not guarantee that the effects due to 
the gravity of the body, such as gravitational radiation and self-force, 
can be neglected when studying the motion of the body.
Issues that arise when including these effects are are discussed briefly
at the end of this paper.

Provided the body can be tossed into the black hole, 
the final composite object would have mass $M+\delta E$, 
angular momentum $J+\delta J$ and charge  $Q+\delta Q$. 
In order for the K-N metric with these parameters to be a naked singularity they would have to satisfy the inequality
\beq
\label{jqover}
(M+\delta E)^2 < \left(\frac{J+\delta J}{M+\delta E}\right)^2+(Q+\delta Q)^2.
\eeq

Various special cases 
of such test body experiments
have been considered in the literature.
Wald focused on an exactly extremal black hole and showed that 
it cannot be overcharged or over-spun using a particle with charge 
and/or orbital angular momentum. He also showed that an extremal
neutral rotating black hole cannot be overspun
using a particle with spin angular momentum falling along the 
spin axis \cite{Wald}. However, de Felice and Yu have shown that 
an extremal charged black hole {\it can} be sent over the extremal limit
by accretion of a neutral spinning test body \cite{deFelice:2001wj}.
If one starts with a non-extremal black hole, 
the results differ from what Wald found for the extremal case.
In particular, Hubeny showed that one can overcharge a {\it near}-extremal 
Reisser--N\"ordstom black hole by tossing in a test body \cite{Hubeny:1998ga},
and Hod showed that a near-extremal rotating black hole can be overspun
in the limiting case where a particle carrying angular momentum is lowered 
all the way to the horizon of a black hole and dropped from there \cite{Hod:2002pm}.
We will return to these results later on and comment on them more extensively.
A case we will not consider further is that of dyonic black holes which
carry both electric and magnetic charge \cite{hiscock,semiz}.
We will also not consider here 
test field experiments
via wave scattering, classical or quantum, as discussed for example in 
\cite{Matsas:2009ww} and references therein.

Astrophysical 
black holes tend to increase their angular momentum by accreting matter, 
whereas they tend to decrease their charge by attracting opposite and 
repelling same charges. For a point of principle, one may 
ignore these facts, but 
nevertheless it would be much more provocative and promising if arguments
showed that a naked singularity could be created using only angular momentum,
since that might then actually occur in nature. 
We therefore focus mostly on the case were the both the black hole 
and the test body have only angular momentum and no charge. 

We shall demonstrate that in the test body approximation a trans-extremal 
condition can be attained, even when taking into account constraints on the
size and structure of the body. The limitations of the test-body approximation
will then be addressed.

\subsection{Over-spinning a neutral black hole}

With $Q=\delta Q=0$, the inequality in eq.~(\ref{jqover})  takes the form
\beq
J+\d J > (M+\d E)^2.
\eeq
This yields a lower bound on the required angular momentum carried by the body, for a given energy $\delta E$:
\beq\label{over-spin}
\d J > \d J_{min}=(M^2-J)+2M \d E+\d E^2.
\eeq
Since we are assuming $\d E\ll M$, it might seem
that the $\d E^2$ term may as well just be neglected
at this stage. However, as we will see 
shortly, 
the presence of that term imposes 
an upper bound on $\d E$ and $\d J$ and, therefore, should not be neglected. 
\\

We can already extract a useful piece of information from 
eq.~(\ref{over-spin}).
Dividing both 
sides
by $\delta E^2$,  
and observing that each term on the right hand side 
is positive and therefore 
should by itself be smaller that the left hand side, we get
\beq\label{largespin}
\d J/\d E^2>2M/\d E \gg1 
\eeq
(where the last inequality follows from (\ref{small}). 
If $\d E$ 
is comparable
to the rest mass of the
body (it can be much less if the body is deeply bound by the gravitational field of the black hole), 
and if $\d J$ comes from spin (rather than orbital angular momentum),
this would imply that the body
has angular momentum far over the 
extremal ratio. 
In that case
the body could not be a black hole. This does not mean 
that it would have to be a naked singularity itself, 
as there is no a priori upper limit to this ratio for bodies other 
than black holes. Stars for instance can easily have ratios much larger than 1.

The requirement that the composite object be a naked singularity 
gave us  
the
lower bound 
(\ref{over-spin})
on the angular momentum of the body.
An upper bound is
obtained from the requirement
that the body does indeed
cross the horizon. 
One can  
use
the equations of motion for the 
body in order to derive the bound~\cite{Wald}. 
These are the Papapetrou equation if the body's angular momentum 
includes
spin, 
and the geodesic equation if it is 
purely
orbital angular momentum. 
But a simpler and more transparent 
method is to just consider the flux of energy and angular momentum 
into the black hole when the body falls across the horizon. The requirement that
the energy momentum tensor satisfies the null energy condition (which follows for
example if the energy density is positive in all local reference frames)
yields 
(see for example \cite{Jacobson:2009kt}) 
 \beq\label{go-in2}
\d E\ge\O_H\d J,
\eeq
where $\O_H=a/2Mr_+$ is the angular velocity of the horizon 
and $r_+=M+(M^2-a^2)^{1/2}$ is the horizon radius
in Boyer-Lindquist 
coordinates. This condition can be written as
 \beq\label{go-in}
\d J \le \d J_{max}=\frac{2Mr_+}{a} \d E.
\eeq
It guarantees that the body can 
fall across the horizon starting from {\it some} point outside,
although in general 
the body is in a bound orbit that does not come 
from spatial infinity.

We now have both an upper and a lower bound for the 
angular momentum of the body, for a given energy. 
As long as $\d J_{min}<\d J_{max}$ for some
$\d E$, there will 
be values of $\d J$ and $\d E$ satisfying both inequalities
(\ref{over-spin}) and (\ref{go-in}). 

First let us 
suppose
the black hole starts exactly extremal, i.e. 
$J=M^2$. Then $a=M=r_+$, so one has
$\d J_{min}=2M\d E +\d E^2$ 
and $\d J_{max}=2M\d E$. 
This implies that $\delta J_{min}> \delta J_{max}$ 
so it is impossible to over-spin
the black hole.
Thus we recover the result of Wald \cite{Wald} mentioned earlier. 
(The analysis here is significantly simpler than that
 of Ref. \cite{Wald}.)
 
The physical interpretation of this result is the following: In the case of the spinning 
body, the spin-spin interaction with the spin of the black hole is sufficiently repulsive
to prevent the body from falling in if it would have overspun the black hole. 
In the orbital angular momentum case, If the body 
has the angular momentum required to overspin 
then 
the impact parameter of the body is too large for it to hit the horizon.

In the sub-extremal case, however,
the inequalities {\it can} be satisfied, as was shown in \cite{Jacobson:2009kt}. 
As mentioned earlier, the limiting case where the body is dropped from a point on the horizon 
had been considered previously by Hod  \cite{Hod:2002pm} 
(it corresponds to  $\d J =\d J_{max}$). 
To understand the range of overspinning parameters, 
it is helpful to visualize the inequalities
graphically. If $\d J_{max}$ and $\d J_{min}$ 
are plotted vs. 
$\d E$, 
the former is a straight line through the
origin, with slope $2M r_+/a$,
while the latter is a parabola with positive
intercept, slope $2M$ at the intercept, and 
curved upwards. 
For a sub-extremal black hole we have $r_+>M>a$, so the slope 
of the  $\d J_{max}$ line is greater than the initial slope of the $\d J_{min}$ parabola. 
Some algebra reveals that 
the parabola always intersects
the straight line in two points. The allowed
values of $\d E$ and $\d J$ are those in the
compact region above the parabola and on or
below the
straight line. Note that
if the $\d E^2$ 
term
is neglected in (\ref{over-spin}),
the parabola is replaced by a straight line,
and 
the allowed region becomes an infinite wedge, with no upper bound.

To 
determine whether the overspinning can be accomplished 
with a small perturbation satisfying (\ref{small}) 
we can expand in the 
dimensionless quantity 
$\e$
defined by 
\beq\label{epsilon}
J/M^2=a/M=1-2\e^2.
\eeq
(Hubeny~\cite{Hubeny:1998ga} used the same parameter to analyze the 
charged case, see below.)
The parameter $\e$ measures how close to extremality the 
black hole is to begin with, .
and we must have $\e\ll1$
if the overspinning perturbation is to be small.
It is now useful to adopt units with $M=1$, 
to keep the expressions simpler. 
Then we 
have, from (\ref{over-spin}) and (\ref{go-in}),  
\bea\label{minmax}
\d J_{min} &=& 2\e^2+2\d E + \d E^2\\
\d J_{max}&=& (2+4\e)\d E,\label{jmax}
\eea
where terms of order $O(\e^2\d E)$ have been dropped
in (\ref{jmax}).
The allowed range of $\d E$ lies where the difference
\beq\label{max-min}
\d J_{max}-\d J_{min} = -2\e^2+4\e\d E - \d E^2 
\eeq
is positive, i.e.
\beq\label{dEe}
(2-\sqrt{2})\e <\d E<(2+\sqrt{2})\e.
\eeq
In particular, $\d E$ must be of order $\e$. 
For a given $\d E$, the allowed values of 
$\d J$ are near $2\d E$, so we must have 
\beq\label{dJ/dE}
\d J\sim \d E.
\eeq

We conclude that if $\e\ll1$ the overspinning values of 
$\d E$ and $\d J$ can indeed be consistent with 
the perturbative requirement.

Note that the {\it width} (\ref{max-min}) of the allowed
range of $\d J$ is only of order $\e^2\ll\e$.
Note also that $a - 1=2\e^2$ is parametrically smaller than $\e$.
For example, if $\e=10^{-2}$, then 
the initial black hole must have 
$a=0.9998$. For a thought experiment,
we can imagine even smaller values of $\e$.\

\subsection{Over-charging or over-spinning a charged black hole}

Let us return to similar conclusions reached in the cases mentioned earlier, where different 
assumptions
for the quantities characterizing the black hole were made.
Hubeny considered   
 the case of adding charge to a charged black hole ($J=\delta J = 0$) \cite{Hubeny:1998ga}. 
In analogy to the spinning case, in the charged case the two constraints are
\begin{align}
&\delta Q>M- Q+\delta E,\\
&\delta Q \le \frac{r_+}{Q}\delta E,
\end{align}
where now $r_+ = M + \sqrt{M^2-Q^2}$.
If the black hole starts out extremal, $M=Q=r_+$, then overcharging is 
impossible. However if $M>Q$, then $r_+>Q$, and one easily 
sees by visualizing the inequalities graphically (now they are both straight lines)
that there is an infinite range of solutions, once $\d E$ is 
greater than a certain minimum value.

de Felice and Yu
considered the case of adding angular momentum to an extremally charged
black hole ($Q=M$, $J=\d Q=0$) ~\cite{deFelice:2001wj}. .
In this case the minimum $\d J$ to overspin
is given by 
\beq\label{deF}
\d J > (M+\d E)\sqrt{2M\d E + \d E^2},
\eeq
and there is no maximum $\d J$, since the only requirement for the
body to fall across the horizon is $\d E\ge 0$, which does not involve $\d J$. 
Note that to lowest order in $\d E/M$ 
the minimum overspinning $\d J$
is $\d J_{\rm min} = M\sqrt{2M\d E}$.

\section{Size and structure requirements}

So far we have characterized the body only by its energy 
$\delta E$, angular momentum
$\delta J$ and charge $\d Q$.
We did not consider restrictions placed on its size and structure:
It should
be sufficiently small to justify use of a 
test particle approximation, and 
it should be composed of matter 
having positive energy density and no 
unphysically large stresses. 
The next step is to take into account these requirements.

\subsection{Adding spin angular momentum to a neutral black hole}

We begin with the 
case of a spinning test body. 
For simplicity we assume that the body 
is dropped along the rotational axis of the black hole.
We first consider the case where the body has $\d E\sim m$,
and is not spinning relativistically, so its spin
angular momentum
is given by $\d J\sim mvR\sim \d E\, v R$, where 
$v$ is the surface velocity and 
$R$ is the equatorial radius. The condition $v<1$ then implies
$R>\d J/\d E$. We saw above that
the ratio $\d J/\d E$ must be of order unity (\ref{dJ/dE}), 
that is
of order $M$. In this case the body must be larger than
the black hole, so it simply will not ``fit" in the transverse
direction, and 
in any case treating it as a point particle with spin
would be unjustified,
since that rests on the
smallness of the size of the body compared to the 
ambient radius of curvature. Moreover, one can 
show \cite{Jacobson:2009kt} 
that the radial tidal 
stress required to hold the body together 
would be larger than the energy density, violating
energy conditions.
It cannot help to allow ultra-relativistic tangential
velocity:  as a simple Newtonian estimate shows, 
that would require unphysical stresses holding the body together.
The conclusion is that it is impossible to 
over-spin the black hole if the body's 
energy is 
close to its rest mass, $\d E\sim m$.

Since the angular momentum involves the
rest mass $m$, not the energy $\delta E$,
it might be possible to achieve a large enough 
$\d J$ with a small enough size $R$, 
without requiring unphysical matter,
by dropping the body from a position
where it is deeply bound, $\d E\ll m$. 
This might be achieved by slowly lowering the
body on a tether,  down to the near the
black hole horizon, before dropping it in.
Now we reconsider whether the size restrictions 
can be met in this setting. 

We begin with the restrictions on the rest mass $m$. 
If $m$ is much greater than $\d E$, then
the test body approximation requires
that we impose not only $\d E\ll 1\, (=M)$ (\ref{small}),
but also $m\ll 1$. There is also a lower bound on $m$,
coming from an upper bound on $R$: 
the angular momentum is  
$\d J\sim mvR$,
hence
(restricting to nonrelativistic spin $v<1$ as required
by the previously mentioned result)
$R > \d J/m\simeq 4\e/m$.
The requirement $R\ll1$ then yields
$m\gg\e$. The mass and size must therefore fall within
the ranges
\beq\label{bounds}
\e\ll m\ll 1, \qquad 4\e/m\lessim R\ll 1.
\eeq
To these conditions we must add the requirement 
$R\gsim m$
that the body is not a black hole, as explained above.

The inequality (\ref{go-in2})
guarantees that the body can cross the horizon with the chosen values
of energy and angular momentum, but since the deeply
bound drop point lies at a finite distance from the horizon it 
is necessary to check that (a) the spinning body would actually fall into the
black hole rather than being repelled, and (b) it is possible to choose 
the polar radius of the
body $R_{\rm polar}$ to be smaller than
the proper distance $d$ from the horizon to the drop point
\beq\label{R<d}
R_{\rm polar}<d,
\eeq
so that it can 
fully ``fit" outside the black hole and be localized at the drop point.
Now it turns out \cite{Jacobson:2009kt} that, in
order to fall in, 
the maximum value that $d$ 
can have, given the allowed values of $\delta E$ and $\delta J$, 
is
\beq
d_{max}\simeq \e/m.
\eeq
Thus we arrive at the bound
\beq
R_{\rm polar}<\e/m.
\eeq
Together with (\ref{bounds}), this means that the body
must be at least somewhat oblate, 
$R_{\rm polar} \lessim R/4$.\footnote{In \cite{Jacobson:2009kt} the
possibility that  $R_{\rm polar} \ne R_{\rm equator}$ was overlooked, 
so it was 
erroneously 
concluded that no value of $R$ could meet all requirements.}
We conclude that 
the body can be large enough
to possess the requisite angular momentum 
and also have
a physically acceptable stress
and 
fit outside the black
hole at the drop point\footnote{de Felice and Yu made a similar
analysis
for 
the case of dropping 
a spinning body into an extremal charged black hole,
but they computed the coordinate radius corresponding to $d_{\rm max}$, 
rather than the proper distance. In the extremal case, the proper distance to the 
horizon is infinite in the direction orthogonal to the Killing vector, so there is
apparently no requirement that the body be disk-shaped, contrary to
what was stated in \cite{deFelice:2001wj}.\label{dFY}}.

\subsection{Adding orbital angular momentum to a neutral black hole}

We 
turn now
to the case
of orbital angular momentum in the equatorial 
plane. Here the issue
is that in order to have the required values of 
$\d E$ and $\d J$, the body might have to be in a bound orbit,
which would have a turning point at a maximum radius.
In that case we would need to require that
the body be small enough to fit outside the horizon at this radius.
Since the body can be no smaller than a black hole
with the same rest mass, it is not clear in advance whether
this requirement could be met. However, as has been shown in \cite{Jacobson:2009kt}
this size constraint is not an issue, since in fact there are 
suitable
orbits that come in from infinity with no turning point. 
This can be shown numerically, but also analytically by the use of 
the effective potential governing the motion of a test particle in a 
Kerr spacetime (Kerr-Newman with no charge). 

\subsection{Cases involving a charged black hole}

Let us briefly 
consider the size and structure requirements
when attempting to overcharge 
or overspin a 
charged 
black hole.
In the case with no angular momentum 
Hubeny~\cite{Hubeny:1998ga} 
showed that the body 
can have the required charge and mass, with 
low internal stresses and size much smaller than
the black hole. Also, she demonstrated that
there
are 
charged
test particle trajectories that fall from infinity into the black hole. 
Therefore, much like the orbital angular momentum case, 
size
constraints are not
an issue. On the contrary, for spinning particles dropped 
radially with radial spin into 
an extremal charged black hole,
de Felice and Yu~\cite{deFelice:2001wj} 
found that the test body must be bound very close to the horizon.
The same is true for a particle carrying orbital angular momentum
but no spin, as we now show.
The radial motion is governed by the 
equation $\dot{r}^2 + (1-1/r)^2(1+\tilde{L}^2/r^2) = \tilde{E}^2$. 
Here $\dot{r}$ is the derivative of the Reissner-Nordstrom radial coordinate
with respect to the particle proper time, and $\tilde{E}=(\d E)/m$ and $\tilde{L}=(\d J)/m$
are the energy and and angular momentum divided by the particle rest mass $m$,
and we have again set $M=1$.
As mentioned after (\ref{deF}), the overspinning requirement is
$\d J/\d E \gsim \sqrt{2/\d E}\gg1$, hence $\tilde{L}/\tilde{E}\gg1$.
For an unbound orbit $\tilde{E}\ge1$, so we infer that $\tilde{L}\gg1$.
There are therefore two turning points where $\dot{r}=0$. 
To fall into the hole the
particle must lie inside the inner turning point,  
which lies at a radius $r_{\rm inner}$ very close to the horizon, 
where $r_{inner}-1\simeq  \tilde{E}/\tilde{L}  \lessim \sqrt{\d E/2}\ll 1$.
However, although the radial coordinate must be very close to that
of the horizon, the proper distance to the horizon, measured in the 
static frame, is infinite for an exactly extremal black hole. 
Hence, in both the spin and orbital cases, 
no further
size constraints 
appear to be imposed by the location
of the turning point (see also footnote
3). 

\section{Beyond the test body approximation}

Our considerations thus far have been
based on an approximation which neglects loss of energy 
and angular momentum in
gravitational  
 radiation and does not take into account self-force effects.  
Recall that 
our purely kinematic considerations above  
yielded a finely tuned relation  between the energy and angular momentum of the 
dropped body for over-spinning to occur.
Both quantities have to be of order $\e$ in
magnitude, but the allowed window for angular momentum, given the 
energy, is only of order $\e^2$. In the light of this delicate balance, it is certainly possible that, 
although small, gravitational radiation and/or 
self-force
effects may always manage to preclude the over-spinning. 

Given that the inequalities (\ref{over-spin}) and (\ref{go-in}) 
need only
hold on the horizon, one could imagine that the loss of energy 
and angular momentum 
in gravitational 
radiation 
might 
be compensated by 
adjusting the initial conditions. 
In the case of an axially symmetric spinning body 
falling along the black hole spin axis
there is no radiation of angular momentum, so it
should be possible to simply compensate for the energy radiated.
To determine whether compensation is actually possible
in the orbital case requires further investigation.   
Perhaps more worrisome than radiation
are
the self force effects.
Indeed
Hubeny found strong indications that for the charged case the 
electromagnetic self-force might  
prevent the overcharging, although her 
calculations were not conclusive \cite{Hubeny:1998ga}. 

Another distinct effect that 
might 
prevent the creation of a naked singularity is 
the tides raised on the black hole horizon by the falling body. 
These would be irrelevant for the orbital angular momentum case
since the body falls in from spatial infinity. 
In the spinning body case, 
however, in which  the body is lowered
to the horizon and then dropped, 
the tidal bulge of the horizon might perhaps 
make it impossible for the body
to start out in the exterior
while still satisfying the size constraints.

Given the 
existing evidence for cosmic censorship, it seems indeed 
likely that neglected 
gravitational effects will come to its rescue. 
The examples discussed here suggest dynamical regimes
in which
it may be interesting to study these neglected effects. 

\section*{Acknowledgements}
This research was supported in part by
the NSF 
under 
Grant Nos. 
PHY-0601800 and PHY-0903572, and by STFC.

\end{document}